\documentclass[journal]{IEEEtran}

\ifCLASSINFOpdf

\else

\fi

\usepackage{makecell}
\usepackage[dvips]{graphicx}
\usepackage{latexsym}
\usepackage{amssymb}
\usepackage{amsmath}
\usepackage{bm}
\usepackage{multirow}
\usepackage{xcolor}
\usepackage{lipsum}
\usepackage{multicol}
\usepackage{enumerate}
\usepackage{algorithm}
\usepackage{algorithmicx}
\usepackage{algpseudocode}
\usepackage{amsmath}
\usepackage{graphicx}
\usepackage{subfig}
\usepackage[utf8]{inputenc}
\usepackage{float}

\begin{document}

\title{1-Bit Massive MIMO Transmission: Embracing Interference with Symbol-Level Precoding}

\author{Ang Li,~\IEEEmembership{Member,~IEEE}, Christos Masouros,~\IEEEmembership{Senior Member,~IEEE}, A. Lee Swindlehurst,~\IEEEmembership{Fellow,~IEEE},\\ and Wei Yu,~\IEEEmembership{Fellow,~IEEE}

\thanks{© 2021 IEEE.  Personal use of this material is permitted.  Permission from IEEE must be obtained for all other uses, in any current or future media, including reprinting/republishing this material for advertising or promotional purposes, creating new collective works, for resale or redistribution to servers or lists, or reuse of any copyrighted component of this work in other works.}
}


\maketitle

\begin{abstract}
The deployment of large-scale antenna arrays for cellular base stations (BSs), termed as `Massive MIMO', has been a key enabler for meeting the ever-increasing capacity requirement for 5G communication systems and beyond. Despite their promising performance, fully-digital massive MIMO systems require a vast amount of hardware components including radio frequency chains, power amplifiers, digital-to-analog converters (DACs), etc., resulting in a huge increase in terms of the total power consumption and hardware costs for cellular BSs. Towards both spectrally-efficient and energy-efficient massive MIMO deployment, a number of hardware limited architectures have been proposed, including hybrid analog-digital structures, constant-envelope transmission, and use of low-resolution DACs. In this paper, we overview the recent interest in improving the error-rate performance of massive MIMO systems deployed with 1-bit DACs through precoding at the symbol level. This line of research goes beyond traditional interference suppression or cancellation techniques by managing interference on a symbol-by-symbol basis. This provides unique opportunities for interference-aware precoding tailored for practical massive MIMO systems. Firstly, we characterize constructive interference (CI) and elaborate on how CI can benefit the 1-bit signal design by exploiting the traditionally undesired multi-user interference as well as the interference from imperfect hardware components. Subsequently, we overview several solutions for 1-bit signal design to illustrate the gains achievable by exploiting CI. Finally, we identify some challenges and future research directions for 1-bit massive MIMO systems that are yet to be explored.
\end{abstract}

\IEEEpeerreviewmaketitle

\section{Introduction}
Massive multiple-input multiple-output (MIMO) has been an attractive technology to support the desired 1000-fold system throughput improvements for fifth-generation (5G) cellular communication systems. By deploying a large-scale antenna array at base stations (BSs) or access points (APs), massive MIMO systems are able to spatially multiplex multiple data streams during the same time and frequency with near-orthogonal transmission simultaneously, offering significant gains in data rates compared to classical small-scale deployments. More importantly, the channel hardening effect further simplifies the channel estimation and power allocation procedure for massive MIMO. Nevertheless, the above potential benefits are premised for ideal fully-digital massive MIMO BSs with infinite-precision digital-to-analog converters (DACs), where losses or signal distortions are not present. In practice, this would require a radio frequency (RF) chain, a highly linear power amplifier (PA) and a pair of DACs per antenna element (higher than 6 bits), and such deployments would result in significant hardware complexity and costs. Furthermore, the vast amount of RF components required by massive MIMO systems significantly increases the total BS power consumption. This is especially true for the downlink, where the PAs and DACs account for the bulk of the power dissipated by the BS. Transmission of signals with a wide dynamic range requires linear PAs with a large back-off that further reduces energy efficiency, while the power consumption increase of the DACs is exponential to the bit resolution and linear to the bandwidth.

\begin{figure*}[!t]
\centering
\includegraphics[width=0.95\textwidth]{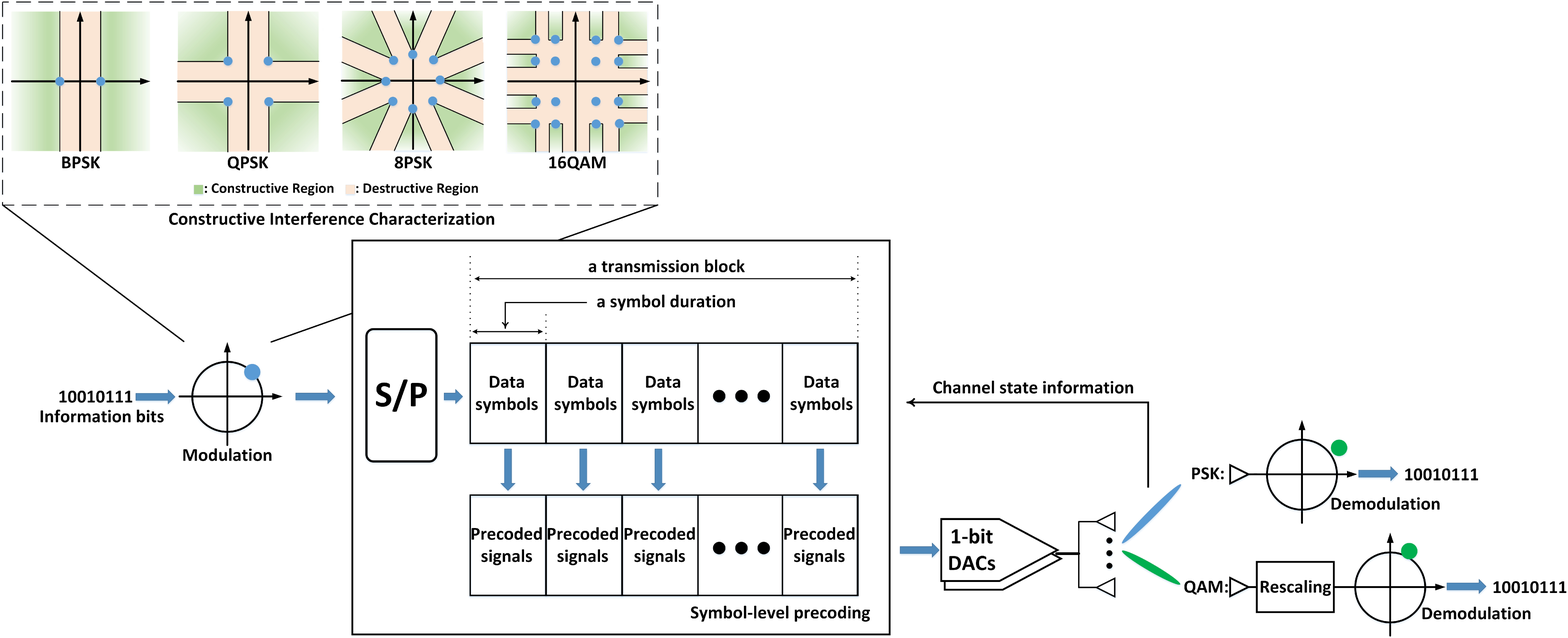}
    \caption*{Figure 1. A generic framework for 1-bit massive MIMO communication systems based on constructive interference.}
\end{figure*}

All of the issues mentioned above have made fully-digital massive MIMO systems prohibitive to deploy in practical wireless communication systems. Towards practical massive MIMO deployment, hardware-efficient massive MIMO architectures have attracted increasing research attention, which can balance the tradeoff between performance, hardware complexity and cost, and power consumption. Existing solutions include analog-only precoding, hybrid analog-digital (HAD) architectures in the millimeter-wave (mmWave) band (defined as FR2 in 5G), constant-envelope (CE) transmission, to name a few \cite{ci-1}. More recently, the use of low-resolution DACs has attracted research attention as a promising solution for hardware-efficient massive MIMO in both the sub-6 GHz band (defined as FR1 in 5G) and the mmWave band, which is the focus of this paper.

Compared to the HAD architecture and CE transmission which reduce the hardware complexity by reducing the total number of RF chains, implementing low-resolution DACs retains the benefits in degrees of freedom for massive MIMO by reducing the hardware cost, complexity and the resulting power consumption on each RF chain \cite{ci-dac-1}. Given that two DACs are required per RF chain, the aforementioned three factors can be greatly alleviated through the implementation of DACs with low resolution, especially the 1-bit DACs. Meanwhile, similar to the case of CE transmission that enjoys low peak-to-average-power-ratio (PAPR) transmission, the 1-bit DAC outputs also meet the CE requirement, which enables to deploy the most power-efficient PAs at each antenna port to improve the energy efficiency. Towards the design of 1-bit precoding, the main difficulty lies in the constraint incurred by the 1-bit DACs, which enforce the transmit signals on each antenna to be selected from the set $\left\{ {\pm 1 \pm j} \right\}$. Such highly non-convex constraints have made it difficult to find the optimal solution for 1-bit precoding. In the following, we overview the precoding approaches for 1-bit DAC transmission based on the mean-squared error (MSE) metric, and further highlight the opportunities and benefits of interference exploitation.

\begin{table*}
\begin{center}
\setlength{\arrayrulewidth}{0.2mm}
\renewcommand{\arraystretch}{1.4}
\scalebox{0.83}
{
\begin{tabular}{| c | c | c | c | c | c | c | }

    \Xhline{2\arrayrulewidth}
    {\bf 1-Bit } & {\bf Design} & \multirow{ 2}{*}{{\bf Methodology}} & {\bf Linear or} & \multirow{ 2}{*}{{\bf Error Rate}} & \multirow{ 2}{*}{{\bf Complexity}} & \multirow{ 2}{*}{{\bf Feature}} \\
    {\bf Precoding Scheme} & {\bf Principle} & & {\bf Nonlinear} & & & \\
    \hline
    \multirow{ 2}{*}{1-bit ZF } & Block-level & \multirow{ 4}{*}{\begin{tabular}{l}  Direct 1-bit\\
                                   quantization \end{tabular}} & \multirow{ 4}{*}{Linear} & \multirow{ 4}{*}{Poor} & \multirow{ 4}{*}{Low} & \multirow{ 4}{*}{Closed-form precoding matrix} \\
    & ZF & & & & & \\
\cline{1-2}
    \multirow{ 2}{*}{1-bit MMSE \cite{dac-MMSE}} & Block-level & & & & & \\
    & MMSE & & & & & \\
    \hline
    \multirow{ 2}{*}{1-bit C1PO \cite{dac-vlsi}} & \multirow{ 12}{*}{\begin{tabular}{l}Symbol-level\\
                                    {\kern 10pt}  MSE \end{tabular}}  & Biconvex relaxation + & \multirow{ 14}{*}{\begin{tabular}{l} {\kern 5pt} Nonlinear\\ (Symbol-level) \end{tabular}} & Good with & \multirow{ 2}{*}{Moderate} & \multirow{ 2}{*}{Applicable to all modulations}  \\
    & & alternating optimization & & PSK signaling  &  & \\
    \cline{1-1}
    \cline{3-3}
    \cline{5-7}
    \multirow{ 2}{*}{1-bit C2PO \cite{dac-vlsi}} & & forward-backward & & Good with & Moderate, less & Strong connection with \\
    && splitting method & & PSK signaling & than C1PO & C1PO via series expansion \\
    \cline{1-1}
    \cline{3-3}
    \cline{5-7}
    \multirow{ 2}{*}{1-bit SDR \cite{dac-nonlinear}} & & Semidefinite programming & & \multirow{ 2}{*}{Near-optimal} & \multirow{ 2}{*}{High} & Only applicable to \\
    & & and relaxation & &  &  & moderately-sized systems \\
    \cline{1-1}
    \cline{3-3}
    \cline{5-7}
    \multirow{ 2}{*}{1-bit SQUID \cite{dac-nonlinear}} & & Douglas-Rachford & & \multirow{ 2}{*}{Promising} & \multirow{ 2}{*}{Moderate} & Comparable performance to \\
    & & splitting & & & & SDR with low complexity \\
    \cline{1-1}
    \cline{3-3}
    \cline{5-7}
    \multirow{ 2}{*}{1-bit SP \cite{dac-nonlinear}} & & Sphere precoding & & \multirow{ 2}{*}{Near-optimal} & \multirow{ 2}{*}{Prohibitive} & Only applicable to \\
    & & + tree search & &  &  & small-sized systems \\
    \cline{1-1}
    \cline{3-3}
    \cline{5-7}
    \multirow{ 2}{*}{1-bit-Greedy \cite{Sohrabi_Liu_Yu}} & & \multirow{ 2}{*}{Greedy + exhaustive search} & & \multirow{ 2}{*}{Near-optimal} & \multirow{ 2}{*}{Moderate} & With proper constellation range design \\
    & & & &  &  & and with large antenna array \\
    \cline{1-3}
    \cline{5-7}
    \multirow{ 2}{*}{1-bit MM \cite{dac-mingjie}} & SER & Penalty method +  & & \multirow{ 2}{*}{Promising} & \multirow{ 2}{*}{Moderate} & QAM only, applicable to 1-bit DACs, \\
    & minimization & majorization-minimization & & & & continuous  and discrete CE transmission \\

\Xhline{2\arrayrulewidth}

    \multirow{ 2}{*}{1-bit BB \cite{dac-bb}} & \multirow{ 9}{*}{CI} & \multirow{ 2}{*}{Full branch-and-bound} & \multirow{ 9}{*}{\begin{tabular}{l} {\kern 5pt} Nonlinear\\ (Symbol-level) \end{tabular}} & \multirow{ 2}{*}{Optimal} & \multirow{ 2}{*}{Prohibitive} & PSK only, only applicable \\
    & & & & & & to small-sized systems\\
    \cline{1-1}
    \cline{3-3}
    \cline{5-7}
    \multirow{ 2}{*}{1-bit LP \cite{ci-dac-lp}} & & Linear programming & & Good with & \multirow{ 2}{*}{Moderate} & Applicable to both \\
     &  & formulation & & PSK signaling & & PSK and QAM \\
    \cline{1-1}
    \cline{3-3}
    \cline{5-7}
    1-bit SS \cite{ci-dac-1} & & 3-stage algorithm & & Good & Moderate & PSK only \\
    \cline{1-1}
    \cline{3-3}
    \cline{5-7}
    \multirow{ 2}{*}{1-bit IST \cite{dac-zeropower}} & & Iterative soft thresholding & & \multirow{ 2}{*}{Promising} & \multirow{ 2}{*}{Moderate} & PSK only, allows \\
    & & + bit flipping & & & & zero-power allocation \\
    \cline{1-1}
    \cline{3-3}
    \cline{5-7}
    \multirow{ 2}{*}{1-bit P-BB \cite{ci-dac-2}} & & \multirow{ 2}{*}{Partial branch-and-bound} & & \multirow{ 2}{*}{Near-optimal} & \multirow{ 2}{*}{Moderate} & Applicable to both \\
    & & & & & & PSK and QAM \\
\Xhline{2\arrayrulewidth}

\end{tabular}
}
\end{center}

\caption{A summary of linear and nonlinear 1-bit precoding schemes in current literature.}
\end{table*} 

\subsection{Classical 1-Bit DAC Transmission: MSE-Based Approaches}
Precoding design for 1-bit DACs ranges from adaptation of traditional linear precoding to optimization-based nonlinear methods. Of these, the simple adaptation of traditional linear zero-forcing (ZF) precoding and MSE-based approaches, which apply a closed-form precoding matrix to a stream of data symbols over time in a transmission block (a block of symbols), within which the wireless channel remains constant, requires the lowest computational complexity. These techniques yield poor error-rate performance due to the severe signal distortions from the 1-bit DACs that cannot be compensated with block-level precoding. \cite{dac-MMSE}.

Providing a reasonable performance with 1-bit DACs necessitates a signal design symbol by symbol, i.e., symbol-level precoding. Different from traditional block-level precoding designs in which precoding matrices depend only on the channel state information (CSI), in such approaches the signals to be transmitted are directly designed, where the knowledge of CSI and data symbols is both exploited. Moreover, given that the precoding matrices for symbol-level precoding may not have explicit expressions and the precoding procedure may not be linear, in the 1-bit precoding literature they are thus referred to as `nonlinear precoding' methods and usually the precoded signals to be transmitted are directly designed, as shown in Figure 1. Targeting the symbol-level MSE minimization between the intended data symbols and received symbols subject to 1-bit constraint, one line of research employs a least-square approach in \cite{dac-vlsi}, where the biconvex relaxation framework is adopted for 1-bit precoding design, and the resulting biConvex 1-bit PrecOding (C1PO) algorithm is shown to be superior to the 1-bit linear approaches. For very large multi-antenna systems, \cite{dac-vlsi} further provides a low-complexity alternative to the 1-bit C1PO algorithm based on forward-backward splitting, referred to as `1-bit C2PO'. Near-optimal 1-bit precoding solutions relying on the semidefinite relaxation (SDR) method as well as the adaptation of sphere decoding have been employed in \cite{dac-nonlinear}, both of which are applicable to small- and moderately-sized systems. For QAM modulation, \cite{dac-mingjie} proposes a 1-bit precoding framework for SER minimization via the combination of a novel penalty method and an inexact majorization-minimization process. In these traditional MSE-based 1-bit precoding solutions, interference has been regarded as a detrimental factor which is minimized using the MSE metric.

While the works mentioned above focus on optimization techniques for finding the appropriate transmit signals for fixed constellations, an equally important dimension in 1-bit precoding is constellation design. The 1-bit precoding problem can be thought of assigning $\{\pm 1 \pm j \}$ to each of the transmit antennas so that for the given channel, a desired constellation point is synthesized at the receiver. It turns out that in the regime of a large-scale transmit antenna array, it can be shown that restricting the transmit signal to be $\{\pm 1 \pm j\}$ as with 1-bit DACs only provides $\sqrt{\frac{2}{\pi}}$ or 80\% of the dynamic range at the user side compared to the case of infinite-resolution transmit signals \cite{Sohrabi_Liu_Yu}. As long as the constellation range is reduced to about 80\% of the infinite-resolution case, the 1-bit precoder design problem becomes considerably easier -- heuristic greedy and exhaustive search algorithms can already do well if the scale of transmit antennas deployed at the cellular BS goes large \cite{Sohrabi_Liu_Yu}.

We refer the readers to Table I which provides a thorough summary of representative 1-bit precoding schemes in current literature. Compared with linear closed-form 1-bit methods, nonlinear 1-bit precoding schemes are shown to offer significant error-rate improvements, but require increased complexity since an iterative optimization problem must be solved at the symbol rate \cite{dac-nonlinear}.

\subsection{1-Bit DAC Transmission: The Scope for Interference Exploitation}
While both the linear and nonlinear 1-bit precoding techniques mentioned above have their distinct benefits and drawbacks, we note an important feature that has been neglected in these 1-bit precoding designs: When we shift the precoding design from the traditional block level to the symbol level, the MSE-based approach shown in Figure 2 that aims to suppress all interference is no longer optimal. As will be explained in the following, interference need not be constrained in all directions around the intended data symbol. Instead, interference exploitation techniques through the characterization of constructive interference (CI) and destructive interference (DI) are able to exploit beneficial interference inherent in multi-user transmission and the interference artificially introduced from 1-bit quantization. This factor was not fully explored in the traditional MSE-based 1-bit precoding solutions mentioned above, where all the interference is minimized, and can yield additional performance improvements \cite{ci-1}, as will be shown in Section III.

With the above motivation, this paper provides an overview of CI-based symbol-level precoding designs tailored for 1-bit massive MIMO systems. We begin by characterizing the notion of CI and explain how it can benefit 1-bit transmission. Subsequently, we discuss the extension of interference exploitation to 1-bit massive MIMO systems by describing several CI-based 1-bit solutions.

\section{CI Characterization}
We begin by introducing the concept of CI, followed by the description of CI signal design for downlink 1-bit massive MIMO.

\subsection{The Concept of Interference Exploitation}
Information theory and the notion of dirty-paper coding indicate that known interference does not degrade the capacity of the wireless broadcast channel when CSI is available, and that it is optimal to use interference-cognizant coding rather than just cancelling it. In addition to CSI, the data to be transmitted is also available at the BSs, but such information is not fully exploited by traditional block-level precoding methods. This is the idea behind CI exploitation, as illustrated in Figure 2 for the $1+j$ constellation point of quadrature phase shift keying (QPSK) modulation. In this case, QPSK's constellation decision boundaries are formed by the real and imaginary axes, and we can define CI as the interference that is able to push the received signal farther away from these axes. Such ``interference'' becomes beneficial to decoding as it increases the power of the useful signal \cite{ci-2}. As a toy example, consider a two-user scenario in which the desired and the interfering data symbols are $s_1=1$ and $s_2=-1$ respectively, both drawn from a BPSK constellation. For simplicity, assume the channel between the transmitter and receiver is $h_1=1$, and the interfering channel is $h_2=\rho$. When $\rho>0$, the received signal $r=h_1s_1+h_2s_2=1-\rho<1$, which means that the received symbol is pushed closer to the decision boundary by the interfering signal, and hence destructive interference is observed. On the contrary, when $\rho<0$, the received signal $r=h_1s_1+h_2s_2=1-\rho>1$, and the received symbol is pushed farther away from the decision boundary by the interfering signal, in which case constructive interference is observed.

Interference exploitation is achieved by controlling the interfering signals' magnitude and phase through symbol-level precoding, and all of the interfering signals can be made constructive to the signal of interest, allowing the received signals to locate inside the constructive area of the constellation instead of the MSE region and move farther from the decision boundaries of the constellation, leading to improved performance. Consequently, the MSE metric that aims to minimize the difference between the received and transmitted symbol as in Figure 2 is sub-optimal. Given that the interference due to 1-bit DACs is a symbol-level phenomenon and that the existing 1-bit precoding solutions mentioned above neglect the use of CI to further benefit the 1-bit precoding, the concept of interference exploitation finds particular application to 1-bit transmission.

\subsection{Signal Design}
The essential idea behind the CI-based signal design is aimed at exploiting interference instead of suppressing it via symbol-level precoding, pushing the received symbols as far away as possible from their corresponding decision boundaries. The resulting increased distance to the decision boundaries translates into an increase in the received SINR, thus leading to an improved performance for each receiver.

\begin{figure}[!t]
\centering
\includegraphics[width=0.45\textwidth]{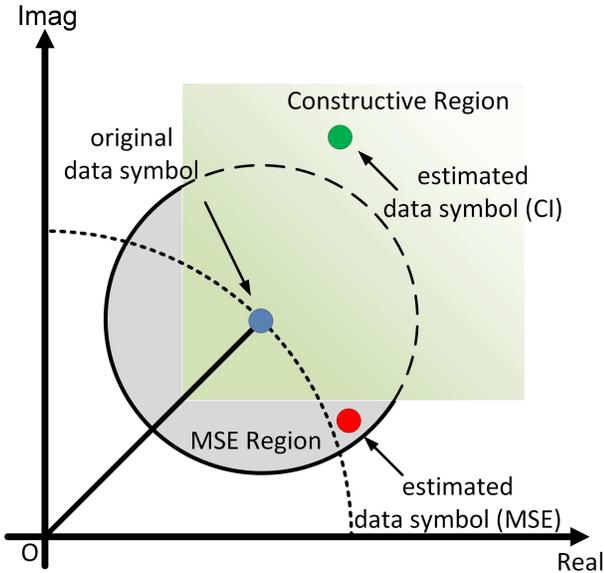}
\caption*{Figure 2. An illustration of the advantage of the CI metric compared with MSE.}
\end{figure}

Recall Figure 1 where we depict the CI-based 1-bit precoding procedure for a generic multi-user transmission scenario. Based on the modulation type, the constructive region within which all the interfered signals achieve CI is determined, depicted as the green shaded areas shown in Figure 1. For example, since the decision boundary for binary phase shift keying (BPSK) modulation is only the imaginary axis by definition, a interfering signal becomes constructive as long as its real part shares the same sign as the real part of the intended signal. Interference can be similarly characterized for higher-order PSK modulations and quadrature amplitude modulation (QAM), and we refer the readers to \cite{ci-1} for more explicit descriptions and mathematical formulations. Subsequently, given that both the data symbols and CSI are available at the BS, symbol-level precoding techniques design the transmit signal vector such that all the interfering signals for each user are simultaneously made constructive to the signal of interest. By exploiting CI, the interfering signals that were harmful to the wireless transmission now become beneficial. Moreover, the transmit power that was used to suppress interference in traditional block-level precoding methods can now be used more judiciously via symbol-level precoding. The above two effects jointly offer a performance gain in terms of signal-to-noise ratio (SNR), which is over 7.5dB for PSK modulations and 5dB for QAM modulations, when compared with traditional ZF and MMSE precoding methods with a $10^{-3}$ uncoded bit error rate (BER) target in a typical small-scale $12 \times 12$ MIMO system \cite{ci-1}.

Similarly, for 1-bit precoding, in addition to the inter-user interference that is inherent in wireless transmission, there also exists artificial interference introduced by the coarse quantization of the 1-bit DACs. In this rich interference environment, both sources of interference can be made constructive to the users, as discussed in the following.

\section{CI-Based 1-Bit Precoding Solutions}
Several CI-based 1-bit solutions have already appeared in the literature of precoding, adopting the CI concept either implicitly or explicitly. For example, \cite{dac-bb} proposes the optimal 1-bit solution through the use of branch-and-bound (BB) framework, which applies to small-sized MIMO systems due to the prohibitive computational costs of the BB operations. In addition, \cite{dac-zeropower} enables zero-power allocation for part of the antenna elements in 1-bit precoding design, and has shown that such an approach enables further performance improvements under certain scenarios. It should be mentioned that although the concept of CI is not mentioned explicitly, the problem formulation in the two works above coincides with the symbol-scaling formulation in \cite{ci-dac-1}. Below, we highlight some representative CI-based 1-bit precoding designs.

\begin{figure*}[!t]
\captionsetup[subfigure]{labelformat=empty}
\begin{centering}
\subfloat[(a) Uncoded BER v.s. SNR, 8PSK, $N_\text{T}=128$, $K=8$]
{\begin{centering}
\includegraphics[width=0.3\textwidth]{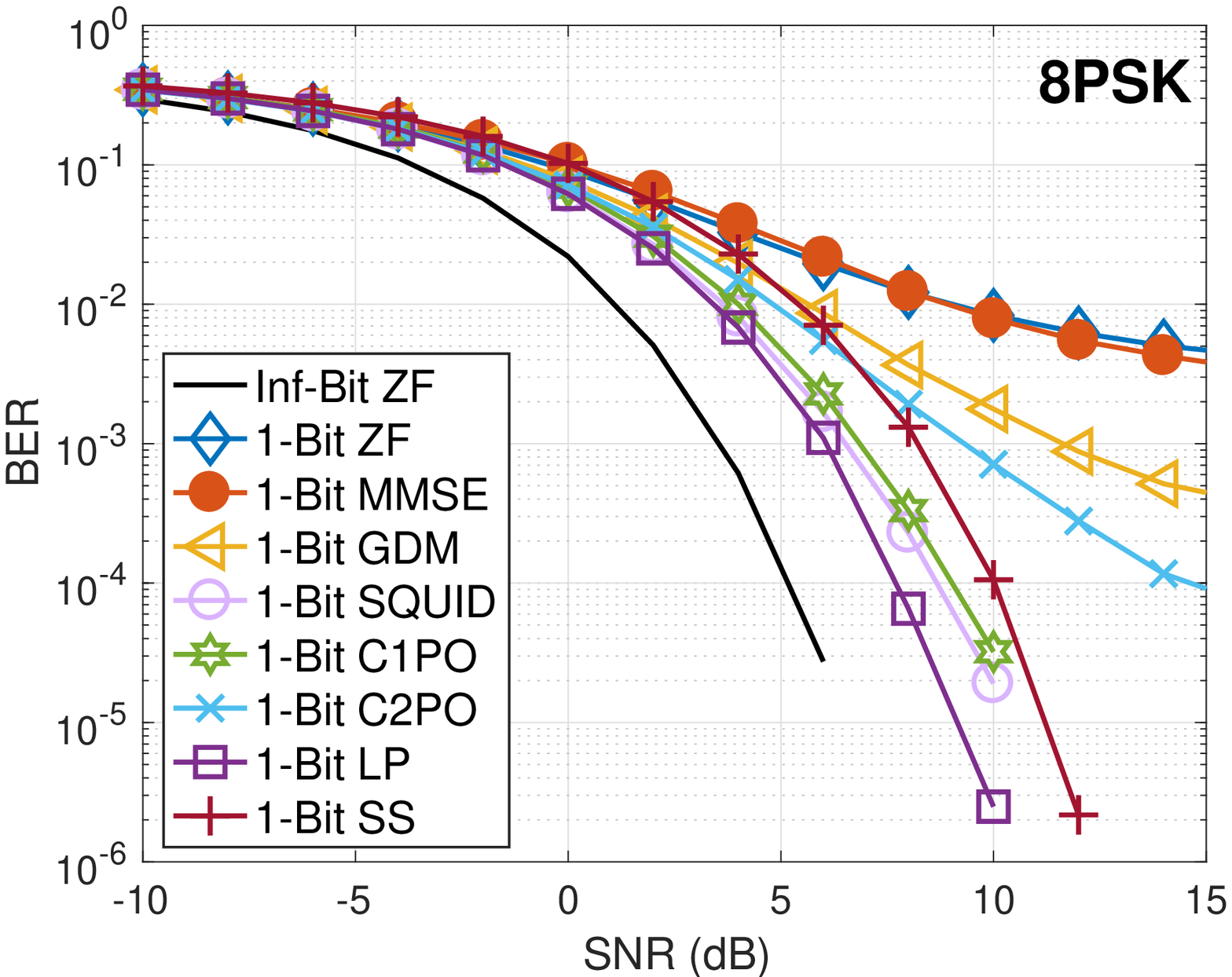}
\par
\end{centering}
}
\hspace{0.2cm}
\subfloat[(b) Uncoded BER v.s. complexity, 8PSK, $K=8$, SNR$=$10dB]
{\begin{centering}
\includegraphics[width=0.3\textwidth]{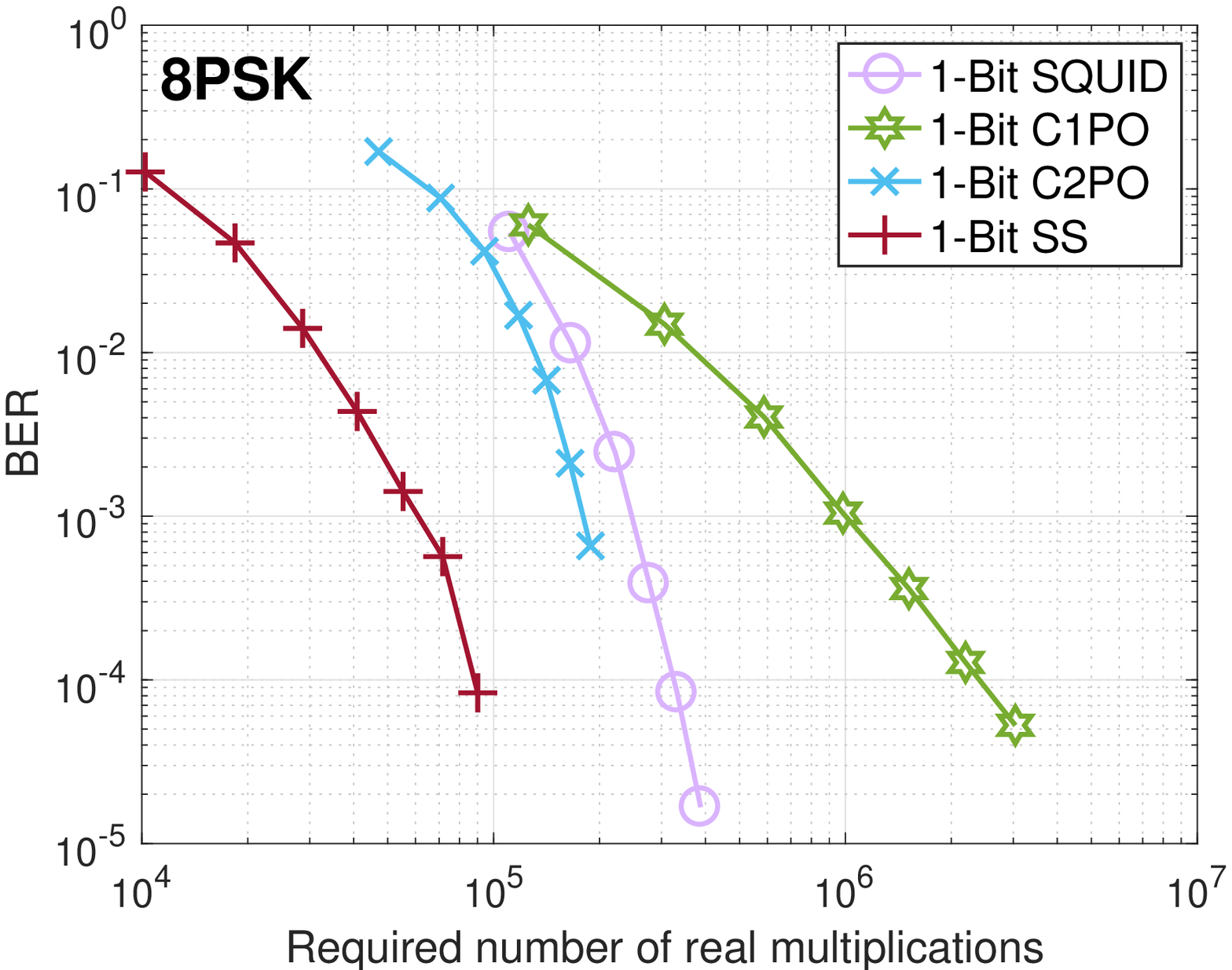}
\par
\end{centering}
}
\hspace{0.2cm}
\subfloat[(c) Uncoded BER v.s. SNR, 64QAM and 256QAM, $N_\text{T}=128$, $K=8$]
{\begin{centering}
\includegraphics[width=0.3\textwidth]{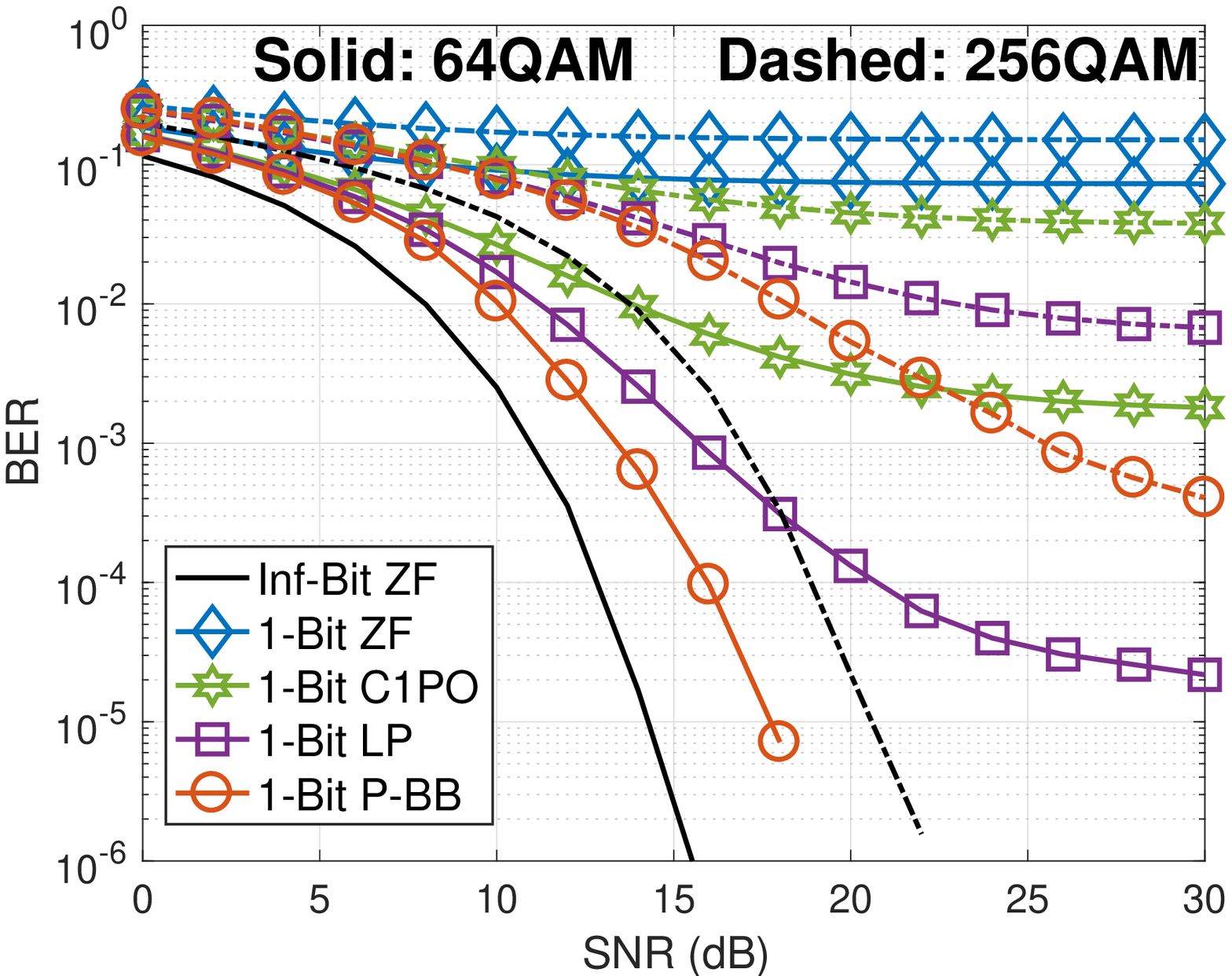}
\par
\end{centering}
}
\par
\end{centering}
\caption*{Figure 3. Numerical comparison of CI-based and MSE-based 1-bit precoding in the literature, standard uncorrelated Rayleigh fading channel, max iteration number for `1-Bit SQUID', `1-Bit C1PO' and `1-Bit C2PO' is 50, 20 and 20, respectively \cite{ci-dac-1}, \cite{ci-dac-2}.}
\end{figure*}

\subsection{Linear Programming Based Solution}
We first introduce a computationally efficient 1-bit precoding solution based on the linear programming (LP) formulation. This approach is formulated in \cite{ci-dac-lp}, and designs the 1-bit precoding which aims to maximize the safety margin (the constructive area defined in the CI concept shown as the green shaded region in Figure 2) to the decision thresholds by incorporating the classical CI constraint into the 1-bit precoding optimization problem. As opposed to traditional CI precoding for fully-digital MIMO systems where the formulated problem is a second-order cone programming (SOCP), a simple LP formulation can be obtained by relaxing the discrete 1-bit constraints to inequality constraints on the real parts and imaginary parts for the transmit signal, and the solution is then obtained by enforcing a 1-bit quantization on the result of the LP optimization. The performance gains of the CI-based 1-bit LP solution includes promising error-rate performance with low computational complexity, obtained by exploiting interference rather than minimizing it. To illustrate the corresponding error-rate performance, Figure 3(a) presents the BER result with an increase in the SNR at the transmitter side in a typical massive MIMO scenario. Standard uncorrelated Rayleigh fading channels are considered, and there are $K=8$ single-antenna users in the scenario with $N_\text{T}=128$ transmit antennas at the BS, benchmarked by several linear and nonlinear 1-bit solutions in current literature. Since DACs are performed in the baseband, 1-bit precoding schemes apply to both the sub-6 GHz and mmWave bands, although standard uncorrelated Rayleigh fading channels are adopted to generate the numerical results in Figure 3 as a representative example. As observed, the `1-Bit LP' \cite{ci-dac-lp} exhibits the best BER result compared with other benchmark schemes and the SNR loss compared to unquantized ZF precoding is less than 3dB when the BER target is $10^{-5}$, bearing in mind that only 1-bit DACs instead of high-resolution ones are employed.

\subsection{Symbol Scaling Based Solution}
We next illustrate an alternative low-complexity 1-bit precoder based on symbol scaling. Conceptually, the symbol scaling performs a decomposition for both the data symbols and the 1-bit transmit signals along the decision boundaries (the real and imaginary axis in Figure 2 for QPSK modulation as an example). The mathematical relationship between these approaches is formulated in \cite{ci-dac-1}, along with a three-stage algorithm based on this idea, referred to here as `1-bit SS'. In particular, the values of the precoded signals for some antennas are allocated in the first stage following a specific criterion based on the data symbol and CSI. The second stage allocates the 1-bit signals for the residual entries of the transmit signals based on two different design criterions. The final stage includes a refinement process which performs a greedy algorithm to see if the CI effect can be further improved by modifying the sign of each precoded signal. The superiority of the `1-bit SS' scheme lies in its performance-complexity tradeoff, achieved using an efficient iterative algorithm instead of solving an optimization problem. To illustrate the favorable performance-complexity tradeoff offered by the proposed 1-bit SS scheme, Figure 3(b) plots the BER against the analytical complexity for $K=8$ users with a transmit SNR of 10dB, where the number of transmit antennas varies from 32 to 128. As can be observed, the 1-bit SS scheme is superior to 1-bit algorithms in the existing literature, requiring only 10\% the complexity at the cost of only slight BER losses.

\subsection{P-BB Based Near-Optimal Solution}
Next we describe a near-optimal 1-bit solution which leverages a partial branch-and-bound (P-BB) framework, applicable to both PSK and QAM modulations. This method is based on that most entries in the 1-bit transmit signal of the solution to the LP formulation discussed in Section III-A already satisfy the 1-bit requirement, and the quantization losses are due to the relatively small portion of entries that fail to obey the 1-bit constraint \cite{ci-dac-2}. The approach in \cite{ci-dac-2} leaves the entries already satisfying the constraint unchanged, and only applies the BB procedure to those for which the constraint is inactive. As a result, the complexity relative to previous BB approaches is significantly reduced, and meanwhile the error-rate performance has been improved over the `1-bit LP' method because the introduced P-BB framework returns a near-optimal solution. To numerically demonstrate the performance gains, in Figure 3(c) we depict the BER result for both 64QAM and 256QAM modulation as the transmit SNR increases, where it is observed that the performance of the traditional interference-suppression approaches degrades severely, although they exhibit competitive error-rate performance when PSK signaling is employed. The P-BB based 1-bit design exhibits an SNR gain of more than 4dB for 64QAM and 6dB for 256QAM over existing 1-bit precoding designs, validating its superiority.

\section{Open Challenges and Future Works}
Hardware-efficient massive MIMO systems based on CI are still an open and ongoing research topic for 5G and future wireless communication systems. Some interesting research directions in the future are identified in the following.

\subsection{CI Precoding for Different DAC Architectures}
One of the potential future works in this field is the study of alternative DAC architectures. A representative work uses a spatial version of Sigma-Delta modulation with 1-bit precoding in \cite{sigma-delta}, which leads to simplified optimization problem formulations for 1-bit precoding by transforming the complicated binary optimization into peak amplitude constraints and better control over the 1-bit quantization for massive MIMO. In addition to the above, using DACs with more than one bit per dimension can be another option, which leads to improved error-rate performance and even approaches the ideal unquantized case, at the expense of increased hardware costs. This problem was considered for the LP approach in \cite{ci-dac-lp} for the case of polar DACs, where the CE property of the transmit signal is maintained by distributing the quantization points around the unit circle. For such studies, a natural question arises: what is the optimal bit resolution for DACs that should be adopted at low, medium, and high SNR regimes regarding the BER, hardware-efficiency and energy-efficiency tradeoff? The literature would benefit from an analytical performance study of massive MIMO with low-resolution DACs, founded on which, practical block/symbol-level few-bit-DAC precoding designs could be developed.

\subsection{Task-Based Quantization}
Recently, a new concept referred to as `task-based quantization (TBQ)' has emerged in the signal processing community as a counterpart to traditional quantization methods \cite{tbq}. The TBQ technique considers the specific task of the system in the quantization design, leading to improved performance compared to traditional quantizers that only aim to accurately represent the underlying signals. The superiority of TBQ over traditional quantization techniques is analogous to that of CI-based 1-bit precoding designs over traditional MSE-based designs, since it takes into account how the signals are processed post-reception. While TBQ has been shown to be a promising technique for channel estimation in the uplink when few-bit analog-to-digital converters (ADCs) are deployed at massive MIMO arrays, such combination of the TBQ concept with interference exploitation has stimulated new research directions for the data transmission of massive MIMO in the downlink.

\subsection{Effect of Nonlinear Power Amplifiers}
Practical wireless communication systems often operate the PAs in the nonlinear region near saturation in order to improve power efficiency, which generally works well with CE transmissions such as CE precoding (CEP) or 1-bit DACs, as described above. However, if few-bit DACs are employed at cellular BSs and the transmit signals are no longer CE, the input signals with increased PAPR may incur severe in-band distortion and out-of-band radiation that further deteriorate the system performance. Thus, the effect of imperfect nonlinear PAs must be taken into account if focus is shifted to higher-order DACs in future works, which would require a rethinking of the communication model that takes into account the resulting signal distortion of the nonlinear PAs. This will potentially lead to a joint optimization problem on the DAC resolution, precoded signals and their corresponding peak power values that need to be carefully designed for a promising performance tradeoff.

\subsection{Machine Learning Based 1-Bit CI Transmitter}
The solutions discussed above for the precoded signal design for 1-bit DACs are all based on model-based approaches, where the intended precoded signal is obtained either by solving an optimization problem or through an (iterative) algorithm, both featuring system parameters that are tuned manually. Recently, machine learning and deep neural networks have extensively been discussed in the literature for their potential applications in wireless communication systems. Machine learning has already found applications in image/audio processing, social behavior analyses, economics and finance, and has been shown to be particularly effective in the field of pattern recognition and natural language processing. The application of machine learning and deep neural networks (DNNs) to 1-bit precoding has already started to emerge \cite{Sohrabi_ICASSP}, especially utilizing the concept of auto-encoders to model the end-to-end 1-bit precoding procedure as a DNN. A key advantage of such an approach is that the channel uncertainty can be taken into account in the training process for the auto-encoder. By using a data-driven approach, neural network based encoder and decoder, along with optimized constellation, can be designed in a manner that makes the overall system robust to the changing channel propagation conditions.

\section{Conclusions}
This article has given an overview of the potential of employing interference exploitation in hardware-constrained massive MIMO systems. We have discussed how interference can be characterized as either constructive or destructive and how interference artificially introduced in downlink 1-bit massive MIMO systems can be manipulated to further benefit the system performance. The overviewed solutions offer prominent gains in terms of the BER performance, validating the potential by exploiting CI to guarantee both energy-efficient and hardware-efficient massive MIMO architectures. Still, there exist a number of open challenges that are yet to be addressed, opening a broad and exciting new research area for the years to come.

\section*{Acknowledgment}
This work was supported by the EPSRC project EP/R007934/1, the Canada Research Chairs Program, the Science and Technology Program of Shaanxi Province under Grant No. 2019KW-007 and No. 2020KW-007, and the Fundamental Research Funds for the Central Universities under Grant xzy012020007.

\ifCLASSOPTIONcaptionsoff
  \newpage
\fi

\bibliographystyle{IEEEtran}
\bibliography{refs.bib}

\begin{IEEEbiography}{Ang Li}
[S'14, M'18] received his Ph.D. degree in Electronic and Electrical Engineering from University College London in 2018. He was a postdoctoral research associate in the University of Sydney. Since March 2020, he has been a Professor in the School of Information and Communications Engineering, Faculty of Electronic and Information Engineering, Xi'an Jiaotong University, Xi'an, China. His research areas are wireless communications, MIMO, precoding, constructive interference, etc.
\end{IEEEbiography}

\begin{IEEEbiography}{Christos Masouros}
[M'06, SM'14] is currently a full Professor in the Department of Electrical and Electronic Engineering, University College London. His research fields include wireless communications, interference mitigation and exploitation, MIMO and multi-carrier communications, communication and radar coexistence. He is an Associate Editor for IEEE Transactions on Communications, IEEE Transactions on Wireless Communications, and the IEEE Open Journal of Signal Processing, and an Editor-at-Large for IEEE Open Journal of the Communications Society. He was an Associate Editor for IEEE Communications Letters, and a Guest Editor for three special issues of the IEEE Journal on Selected Topics in Signal Processing.
\end{IEEEbiography}

\begin{IEEEbiography}{A. Lee Swindlehurst}
[S'83, M'84, SM'99, F'04] is currently a full Professor in the University of California Irvine (UCI), Irvine, US. He was a former Hans Fischer Senior Fellow in the Institute for Advanced Studies at the Technical University of Munich, and a former Associate Dean for Research and Graduate Studies in the Henry Samueli School of Engineering at UCI. He is the past Editor-in-Chief of IEEE Journal on Selected Topics in Signal Processing, and past member of the Editorial Boards for the EURASIP Journal on Wireless Communications and Networking, IEEE Signal Processing Magazine, and IEEE Transactions on Signal Processing.
\end{IEEEbiography}

\begin{IEEEbiography}{Wei Yu}
[S'97, M'02, SM'08, F'14] is now a full Professor in the University of Toronto, Toronto, Canada, and holds a Canada Research Chair (Tier 1) in Information Theory and Wireless Communications. His main research interests include information theory, optimization, wireless communications, and broadband access networks. He is a Fellow of the Canadian Academy of Engineering. He currently serves as the First Vice President of the IEEE Information Theory Society in 2020, and is an Area Editor for the IEEE Transactions on Wireless Communications (2017-2020).
\end{IEEEbiography}

\end{document}